\documentclass[aps,prl,reprint]{revtex4-2} 

\usepackage{float} 
\usepackage{subfigure}
\usepackage[]{caption2} 

\usepackage[utf8]{inputenc}
\usepackage{graphicx}
\usepackage{CJK}
\usepackage{amsmath}
\usepackage{color}
\usepackage{braket}
\usepackage{multirow}
\usepackage{makecell}

\begin{document}
	\title{The Aharonov Casher phase of a bipartite entanglement pair traversing a quantum square ring}
	\author{Che-Chun Huang}
	\affiliation{Department of Physics, National Taiwan University, Taiwan}
	\author{Seng Ghee Tan}
	\email[Corresponding author: ]{csy16@ulive.pccu.edu.tw ; tansengghee@gmail.com}
	\affiliation{Department of Optoelectric Physics, Chinese Culture University, 55 Hwa-Kang Road, Yang-Ming-Shan, Taipei 11114, Taiwan}
	\author{Ching-Ray Chang}
	\affiliation{Quantum Information Center, Chung Yuan Christian University, Taiwan}
	\affiliation{Department of Physics, National Taiwan University, Taiwan}
	\begin{abstract}
		We propose in this article a quantum square ring that conveniently generates, annihilates and distills the Aharonov Casher phase with the aid of entanglement. The non-Abelian phase is carried by a pair of spin-entangled particles traversing the square ring. At maximal entanglement, dynamic phases are eliminated from the ring and geometric phases are generated in discrete values. By contrast, at partial to no entanglement, both geometric and dynamic phases take on discrete or locally continuous values depending only on the wavelength and the ring size. We have shown that entanglement in a non-Abelian system could greatly simplify future experimental efforts revolving around the studies of geometric phases. 
	\end{abstract}
	\maketitle
	\section{Introduction}
	The quantum ring is a useful apparatus to study the physics of electron phase accumulating and interfering over the confined trajectories as prescribed by the design of the ring. Following the successful measurement of the Aharonov Bohm \cite{1} phase, hot on the heels were a slew of experiments that had demonstrated the Aharonov Casher \cite{2} and the Berry-Pancharatnam \cite{3,three} phases.In modern context, the Aharonov Casher phase associates primarily with the spin orbit coupling, particularly in 2D condensed matter systems. Ring structure carved out of a 2D spin-orbital semiconductor to enclose a magnetic field at the center \cite{4,5,6} was proposed to study the simultainety of the Aharonov Casher and the Aharonov Bohm effects on interference. Efforts have also been made to study the parametric effects \cite{7,8,9} of e.g. the Rashba constant, and the time-dependent magnetic field. At around the same time, the Aharonov Casher effect was experimentally measured in a number of ring structures \cite{10,11}. On a separate study, the Aharonov Casher phase is also associated with the non-Abelian gauge field for its spin phase \cite{12,13,14} and spin force effects \cite{15,16,17,seventeen}, categorically reviewed in Ref \cite{18}. While the spin phase which comprises the geometric and the dynamic parts has largely been determined in ring structures, the exact nature of the accumulated phases in these devices remain ambiguous. The dynamic phase remains an elusive component in most cases, and the process to extract the geometric phase continues to be complicated. For example, in Ref \cite{11}, the system is a Rashba 2D that comprises a hedgehog orientation of the effective magnetic fields turned crown-like by a vertical magnetic field. While the strength of the BP phase is proportional to the solid angle subtended in the rest frame of the electron, a dynamic phase proportional to $\sin\theta$ is also formed in concomitance. By applying an in-plane B field, which modifies the geometric Berry-Pancharatnam, and keeps the dynamic unaffected to the first order, a distinction can be made about the two phases. Therefore, isolating the geometric phase is a complicated effort, the Aharonov Casher remains largely a total phase for most applications.\par
\begin{figure}
		\centering
		\includegraphics[scale=0.5]{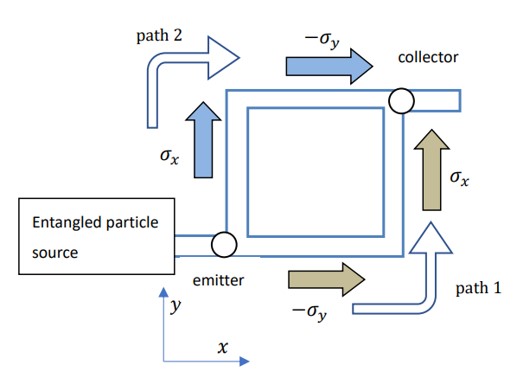}
		\caption{A square quantum ring device that takes a bipartite spin-pair at the emitter and generates an AC total phase as well as its separable components of geometric and dynamics phases.}
		\label{fig1}
	\end{figure}

 
	In this article, we propose a quantum square ring (QSR) that conveniently generates, annihilates or distills the Aharonov Casher phase with the aid of entanglement as shown in FIG. \ref{fig1}. The Aharonov Casher phase generated in this manner comprises the dynamic and the geometric components that can be further separated by tuning the entanglement strength and the device size measured by the wavelength multiple of a traversing particle pair. For example, at maximal entanglement, dynamic phases are eliminated from the device and geometric phases are generated in discrete values. Discrete geometric phases would in turn switch their values on different ring locations depending on the device size. At partial to no entanglement, the Aharonov Casher as well as its dynamic and geometric components can be tuned according to the quantum ring size to take on discrete values or vary continuously across the device. The device is made out of semiconductor or metallic materials that exhibit 2D spin-orbit effects, e.g., the Rashba-Vasko, Dresselhaus, or Dresselhaus-Perel effects \cite{19-1,19-2,19-3,19-4,20-1,20-2,21}. The spin-orbit effects will be the source of both the geometric and the dynamic phases in our system. As the external magnetic field is not needed to generate the geometric phase, nor is it needed to help to eliminate the dynamic phase, a leaner QSR concept that rules out the Aharonov Bohm and the Altshuler–Aronov–Spivak (AAS) effect, and co-opts only the electrically-controllable Aharonov Casher is employed in our design. In the absence of strong B or M field, the adiabatic Berry-Pancharatnam phases in the QSR \cite{22,23-1,23-2,24} is also ruled out. Novel to the functioning of our device though is the entanglement physics \cite{26,27}. On the bottom left of the device is an emitter electrode (FIG. \ref{fig1}) through which an entangled bipartite spin-pair is injected into the QSR. The top right is the collector electrode where the injected spin-pair meets again and carries with it a total phase moderated by the physics of entanglement and device geometry. Our QSR device is therefore, by essence a non-adiabatic and a non-Abelian Aharonov Casher system \cite{28}. The spin-pair’s total phase is accumulated via spin precession about the spin-orbit field but under the constant purview of bi-partite entanglement, which provides in this paper a viable method to generate geometric and dynamic phases in a controllable manner. As an aside, we note that quantum ring device has previously been studied for the practical purpose of producing spin entanglement in a controllable manner. \cite{29,30} There is, however, no discussion on its applicability in the context of geometric phases, let alone any specific discussion on its moderation of the dynamic phases or its distillation of the geometric phases.\par
	
	When a pure quantum state $\ket{\psi(t)}$ evolves on the Hilbert space trajectory in time range $\Gamma:t\in[0,\tau]$, the total phase it accumulates is given by $arg(\langle \psi(0) | \psi(t) \rangle)$. The dynamic phase can be derived from $D = -i\int_{0}^{\tau}\mathrm{d}t \langle \psi(t) | \dot{\psi(t)} \rangle$. One can then define the geometric phase as the result of a total phase minus the dynamic phase as follows
	\begin{align}
		\gamma = arg(\langle \psi(0) | \psi(\tau) \rangle) + i\int_{0}^{\tau}\mathrm{d}t \langle \psi(t) | \dot{\psi(t)} \rangle\label{eq1}
	\end{align}
	Consider a QSR ring of size $\eta \times \eta$ as shown in FIG. \ref{fig1}. The device comprises 2 paths, each consisting of a horizontal and a vertical arm. From the material point of view, the device exhibits the Rashba spin-orbit effect as follows
	\begin{align}
		H = \sigma^{x}k_{y} - \sigma^{y}k_{x}\label{eq2}
	\end{align}
	The QSR geometry conspires with the Rashba effect to generate phase factors for any particle traveling along path 1 and path 2 as follows
	\begin{align}
		U_{I} = e^{\frac{i\delta\sigma_{x}}{2}}e^{\frac{-i\eta\sigma_{y}}{2}}, U_{II} = e^{\frac{-i\delta\sigma_{y}}{2}}e^{\frac{i\eta\sigma_{x}}{2}}\label{eq3}
	\end{align}
	Note that $\delta = \omega t$ for both paths as a result of $\omega_{I} = \omega_{II} = \omega$. Therefore, a spin particle traversing the horizontal arm of Path 1 and the vertical arm of Path 2 would separately accumulate a total phase as denoted by the dimensionless $\eta$. The phase accumulated over time can be translated to a phase at an actual location in space depending on the particle velocity $(\nu)$ in the actual system as denoted by $\omega = k\nu$, where $k$ is the wave-vector. While $\eta$ is hence a phase parameter for the first half of either Path 1 or 2, $\delta\in[0,\eta]$ would represent the phase at any point on the second half of either path. For ease of illustrations, we will refer to the $\eta$ parts of Paths 1 and 2 as, respectively, $\eta_{1}$ and $\eta_{2}$. Likewise, the same is prescribed for $\delta_{1}$ and $\delta_{2}$. The spin-orbit effect when viewed in the rest frame of the carrier is a form of effective magnetic field which sets up a perfect environment for spin precession. As the entangled spin-pair traverses both paths, its phase evolves as prescribed by the unitary operation of
	\begin{align}
		\mathcal{U}\equiv U_{I}\bigotimes U_{II}\label{eq4}
	\end{align}
	The geometric phase would thus be
	\begin{align}
		\gamma = arg(\langle \psi(0) |\mathcal{U}| \psi(0) \rangle) + i\int_{0}^{\tau}\mathrm{d}t \langle \psi(0) |\mathcal{U}^{\dagger}\dot{\mathcal{U}}|\psi(0) \rangle \label{eq5}
	\end{align}
	Explicitly, the dynamic phase is given by
	\begin{equation}
		\begin{split}
			i\mathcal{U}^{\dagger}\dot{\mathcal{U}} = (&I\bigotimes(e^{-\frac{i\sigma_{x}}{2}\eta}(\frac{\sigma_{y}}{2})e^{\frac{i\sigma_{x}}{2}\eta})\\ &- (e^{\frac{i\sigma_{y}}{2}\eta}(-\frac{\sigma_{x}}{2})e^{-\frac{i\sigma_{y}}{2}\eta})\bigotimes I) \\
			=&\frac{1}{2}
			\left(\begin{array}{cccc}
				0 & -i\cos \eta & -\cos \eta & 0 \\
				i\cos\eta & -2\sin\eta & 0 & -\cos\eta \\
				-\cos\eta & 0 & 2\sin\eta & -i\cos\eta \\
				0 & -\cos\eta & i\cos\eta & 0\end{array}\right)
			\label{eq6}
		\end{split}
	\end{equation}
	
	The dynamic phase is expressed in terms of the Pauli matrices so that the relatable picture of effective magnetic fields is not lost. For calculation though, use is often made of its 4 by 4 matrix representation.
	
	\section{Bipartite entangled states}
	The initial states of the entangled-spin-pair at the emitter is then prepared in the Bell-basis of $\ket{\phi(0)}$ or $\ket{\psi(0)}$ as follows:
	\begin{equation}
		\left.\begin{array}{clr}
			\ket{\phi(0)}=\sqrt{p_{0}}\ket{00}\pm\sqrt{p_{1}}\ket{11} \\
			\ket{\psi(0)}=\pm\sqrt{p_{0}}\ket{10}+\sqrt{p_{1}}\ket{01}\end{array}\right\}
		\label{eq7}
	\end{equation}
	where $p_{0}$ and $p_{1}$ determine the strength of the entanglement, and $p_{0},p_{1}\geq0$, $p_{0}+p_{1}=1$.\par
	We will now consider the initial states of $\ket{\phi(0)}=\sqrt{p_{0}}\ket{00}\pm\sqrt{p_{1}}\ket{11}$ to be injected into the QSR through the emitter. As shown in FIG. \ref{fig1}, spin particles 1 and 2 take to paths of their respective namesakes. The geometric phase is the total Aharonov-Casher phase of the system minus the dynamic phase as shown below
	\begin{align}
		\gamma = arg(\cos^{2}(\frac{\delta}{2})\cos^{2}(\frac{\eta}{2})+\sin^{2}(\frac{\delta}{2})\sin^{2}(\frac{\eta}{2})) - D 
		\label{eq8}
	\end{align}
	
	The initial states $\ket{\phi(0)}$ simply could not generate any dynamic phase anywhere on the QSR, i.e. $D=0$. And, the argument of the total Aharonov Casher phase factor consists of parameters that are all real and non-negative. The geometric phase by virtue of $\gamma\equiv arg(a+ib)-D$ vanishes as given by
	\begin{align}
		\gamma = \tan^{-1}\frac{0}{(a>0)}-D\rightarrow \gamma=0\label{eq9}
	\end{align}
	It is clear that the strength of entanglement has no bearing on the geometric and the dynamic phases as $p_{0},p_{1}$ could take on values of the un-entangled states. The results of zero phases might not seem as trivial though. It is a testament to the non-Abelian feature of the QSR device. In fact, the above shows that a bipartite state composed out of $\ket{00}$ and $\ket{11}$ is ideal for eliminating both geometric and dynamic phases from the propagating particles. In terms of applications, this could be a handy device to remove phases where they are not desired from all the particles. In the following, we provide an insight of how dynamic phases are removed from the bipartite state. Let’s examine the dynamic phase by inspecting the constituent particles of the entangled pair. Spin 1 travels on arm-$\eta_{1}$ as though it is in a superposition state of $\ket{0}$ and $\ket{1}$ as far as the dynamic phase is concerned. In either state, its expectation energy is zero as can be deduced by the circular fashion of its spin rotation about the effective magnetic field of $-B_{y}$. Consider spin to precess about the effective magnetic field in an anti-clockwise manner, and that spin 1 to have rotated an angle $\theta<\pi$ by the end of its journey on arm-$\eta_{1}$. Spin 1 would thus continue on arm-$\delta_{1}$ about $+B_{x}$, now inscribing a conical spin rotation with negative energy for $p_{0}$, and positive energy for $p_{1}$. Likewise for spin 2, a corresponding process happens over arm-$\eta_{2}$ about $+B_{x}$ with zero energy for both components $p_{0},p_{1}$. Spin 2 would continue its journey on arm-$\delta_{2}$ about $-B_{y}$, inscribing a conical rotation with positive energy for $p_{0}$, and negative energy for $p_{1}$. The cone energies on arm-$\delta$ cancel one another identically independent of the strength of $p_{0},p_{1}$. The effect is thus a complete negation and a net zero of dynamic phase at all times.\par
	\begin{figure}
		\centering
		\includegraphics[scale=0.5]{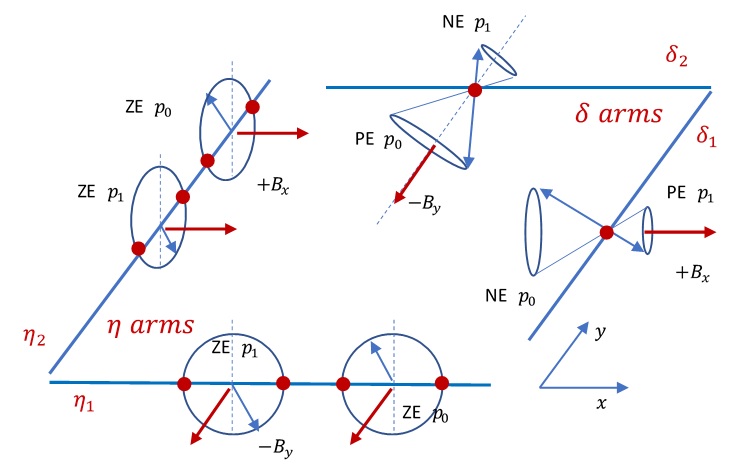}
		\caption{Schematic illustration of the dynamic phases of the bipartite spin pair $\ket{\psi(0)}=\sqrt{p_{0}}\ket{00}\pm\sqrt{p_{1}}\ket{11}$ traversing the QSR. PE, NE, ZE, stand for positive energy, negative energy, zero energy, respectively.}
		\label{fig2}
	\end{figure}
	Note that the energy cones are drawn in different sizes to reflect the energy it carries. This stands against the reality that spin vector is constant in length. Therefore, the energy cones are crude illustrations meant only to provide an intuitive description of the dynamic phases. The precise description of the cone energies is given by the expressions below. Equations (10) and (11) describe the expectation energy for spin particle 1 on arm-$\delta_{1}$.
	\begin{align}
		p_{0}\langle 00|(e^{\frac{i\sigma_{y}}{2}\eta}(-\frac{\sigma_{x}}{2})e^{-\frac{i\sigma_{y}}{2}\eta})\bigotimes I)|00\rangle = -p_{0}\frac{\sin\eta}{2}\label{eq10}
	\end{align}
	\begin{align}
		p_{1}\langle 11|(e^{\frac{i\sigma_{y}}{2}\eta}(-\frac{\sigma_{x}}{2})e^{-\frac{i\sigma_{y}}{2}\eta})\bigotimes I)|11\rangle = p_{1}\frac{\sin\eta}{2}\label{eq11}
	\end{align}
	Equations (12) and (13) describe the expectation energy for spin particle 2 travelling on arm-$\delta_{2}$.
	\begin{align}
		p_{0}\langle 00|(I\bigotimes(e^{-\frac{i\sigma_{x}}{2}\eta}(\frac{\sigma_{y}}{2})e^{\frac{i\sigma_{x}}{2}\eta})|00\rangle = p_{0}\frac{\sin\eta}{2}\label{eq12}
	\end{align}
	\begin{align}
		p_{1}\langle 11|(I\bigotimes(e^{-\frac{i\sigma_{x}}{2}\eta}(\frac{\sigma_{y}}{2})e^{\frac{i\sigma_{x}}{2}\eta})|11\rangle = -p_{1}\frac{\sin\eta}{2}\label{eq13}
	\end{align}
	
	The equations above lend clarity and mathematical credence to our qualitative accounts that the cone energies on arms-$\delta$ cancel one another identically independent of the strength of $p_{0},p_{1}$, resulting in a net zero dynamic phase at all times.\par
	We will now consider the initial states of $\ket{\psi(0)}=\pm\sqrt{p_{0}}\ket{10}+\sqrt{p_{1}}\ket{01}$ to be injected into the QSR through the emitter, once again with the spin particles taking to paths of their respective namesakes. The geometric phase is given by
	\begin{align}
		\begin{split}
			\gamma = arg(\frac{\cos\eta+\cos\delta}{2}&\mp \sqrt{p_{0}p_{1}}(\frac{\sin\delta\sin\eta}{2}) \\ +i(p_{0}-p_{1})&(\frac{\sin\delta\sin\eta}{2})) \\
			+2(\sin\eta)&(p_{0}-p_{1})\delta\label{eq14}
		\end{split}
	\end{align}
	where the dynamic phase is now $-2(\sin\eta)(p_{0}-p_{1})\delta$ and the total phase is deduced accordingly. It is clear from the above that the physics of entanglement has entered the geometric phase. We will study the dynamic phase first. At maximum entanglement where $p_{0}=p_{1}=\frac{1}{2}$, the dynamic phase vanishes with the equality of $p_{0}$ and $p_{1}$, leading to a total Aharonov Casher phase that is purely geometric. Intuitively, at maximal entanglement, the expected energy of the spin-pair is constantly zero. On arm-$\eta_{1}$, inital spin $\ket{1}$ or $\ket{0}$ would precess about an effective $-B_{y}$ field in a circular fashion, both with a zero expectation energy. Likewise on arm-$\eta_{2}$, due to entanglement, the corresponding spin of $\ket{0}$ or $\ket{1}$ would precess about an effective $+B_{x}$ field in a circular fashion, and once again both with a zero energy. In short, circular rotation translates to zero expectation of the Zeeman energy on both arms. Therefore, regardless of entanglement strength, dynamic phase is zero anywhere on arms-$\eta$. The $\delta$ sections of the QSR would, however, generate a dynamic phase at any strength of entanglement other than the maximum, i.e. when $p_{0}\neq p_{1}$. This is because the initial states on arms-$\delta$ are determined by the duration of precession on arms-$\eta$. 
	\begin{figure}
		\centering
		\includegraphics[scale=0.5]{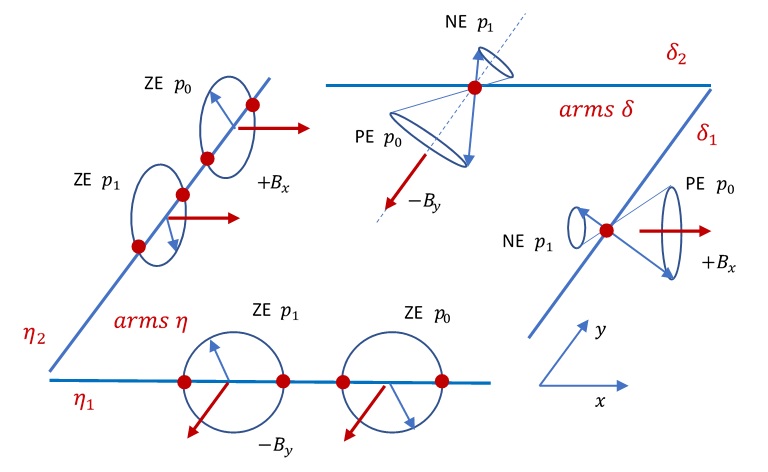}
		\caption{Schematic illustration of the dynamic phases of the bipartite spin pair $\ket{\psi(0)}=\pm\sqrt{p_{0}}\ket{10}+\sqrt{p_{1}}\ket{01}$ traversing the QSR. PE, NE, ZE, stand for positive energy, negative energy, zero energy, respectively.}
		\label{fig3}
	\end{figure}
	On path 1, let spin rotates an angle $\theta<\pi$ about $-B_{y}$ by the end of arm-$\eta_{1}$. Spin 1 would continue on arm-$\delta_{1}$ about $+B_{x}$,  inscribing a conical spin rotation with positive energy for component $p_{0}$, and negative energy for component $p_{1}$. Likewise for path 2, a corresponding process that happens over $\eta_{2}$ about $+B_{x}$ would continue on arm-$\delta_{2}$ about $-B_{y}$, inscribing once again a conical rotation with positive energy for $p_{0}$, and negative energy for $p_{1}$, as shown in FIG.\ref{fig3} It is clear that, on arms-$\delta$, component $p_{1}$ presents a counter effect of proportion $p_{1}$ to the energy due to $p_{0}$. The effect is thus a complete negation and a net zero dynamic phase on the equality of $p_{0}=p_{1}$. The energy cones are drawn in different sizes to reflect the energy it carries. This clearly stands against the quantum reality that spin vector is constant in length. Therefore, the energy cones are crude illustrations meant only to provide an intuitive description of the dynamic phases. The precise description of the cone energies is given by the expressions below. Equations (15) and (16) describe the expectation energy for spin particle 1 on arm-$\delta_{1}$.
	\begin{align}
		p_{0}\langle 10|(e^{\frac{i\sigma_{y}}{2}\eta}(-\frac{\sigma_{x}}{2})e^{-\frac{i\sigma_{y}}{2}\eta})\bigotimes I)|10\rangle = p_{0}\frac{\sin\eta}{2}\label{eq15}
	\end{align}
	\begin{align}
		p_{1}\langle 01|(e^{\frac{i\sigma_{y}}{2}\eta}(-\frac{\sigma_{x}}{2})e^{-\frac{i\sigma_{y}}{2}\eta})\bigotimes I)|01\rangle = -p_{1}\frac{\sin\eta}{2}\label{eq16}
	\end{align}
	Equations (17) and (18) describe the expectation energy for spin particle 2 on arm-$\delta_{2}$.
	\begin{align}
		p_{0}\langle 10|(I\bigotimes(e^{-\frac{i\sigma_{x}}{2}\eta}(\frac{\sigma_{y}}{2})e^{\frac{i\sigma_{x}}{2}\eta})|10\rangle = p_{0}\frac{\sin\eta}{2}\label{eq17}
	\end{align}
	\begin{align}
		p_{1}\langle 01|(I\bigotimes(e^{-\frac{i\sigma_{x}}{2}\eta}(\frac{\sigma_{y}}{2})e^{\frac{i\sigma_{x}}{2}\eta})|01\rangle = -p_{1}\frac{\sin\eta}{2}\label{eq18}
	\end{align}
	Spin particles on arms $\delta_{1}$ and $\delta_{2}$ reinforces once another, the $p_{0}$ and $p_{1}$ components become more positive and negative, respectively. As the equality of the entanglement strength is crucial for suppressing the dynamic phase on arms-$\delta$ but not on arms-$\eta$, a net dynamic phase would accumulate on arms-$\delta$ on the condition of $p_{0}\neq p_{1}$. There is, however, an exception. If the length of arms-$\eta$ translate to a spin rotation of $\eta = n\pi$, subsequent conical precession on arms-$\delta$ would not have happened. Spin would simply continue with circular precession and a zero dynamic phase throughout. The physics above lends further credence to the applicability of the QSR design as a phase purifier. Tuning the entanglement strength to $p_{0}=p_{1}$ at the source, a maximally-entangled spin-pair injected at the emitter would propagate with a dynamic phase suppressed throughout. In the event of $p_{0}\neq p_{1}$ though, dynamic phase could be suppressed by choosing the length of arms $\eta = n\pi$. Having completed our study of the dynamic phase, we will now examine the geometric phase, $\gamma$. At maximal entanglement, i.e. $p_{0}=p_{1}=\frac{1}{2}$, the imaginary part of $\gamma$, denoted by $b$ as shown in Equation (19) below vanishes. The geometric phase is either $0$ or $\pi$ depending on the parameters of the real part $a(\delta,\eta)$ as shown in the denominator of $\tan\gamma=\frac{b}{a(\delta,\eta)}$, where a positive denominator corresponds to $\gamma=0$ while a negative denominator corresponds to $\gamma=\pi$.
	\begin{align}
		\gamma = arg(\frac{1}{2}(\cos\delta+\cos\eta)-\frac{1}{4}\sin\delta\eta+i0)\equiv arg(a+ib)\label{eq19}
	\end{align}
	Let us now study in slightly more details the geometric phase of the spin-pair traversing the $\delta$ arms. The crucial quantity here is range $0<\delta\leq\eta$. In the case of no entanglement, $(p_{0},p_{1})=(0,1)$ or $(1,0)$
	\begin{align}
		\gamma = \tan^{-1}{\frac{\mp(\sin\delta\sin\eta)}{\cos\delta+\cos\eta}}\mp2(\sin\eta)\delta\label{eq20}
	\end{align}
	Dynamic phase is eliminated at arms’ length corresponding to $\eta=n\pi$. Note that when $\delta>0$, phase $\eta$ corresponds to the end location of arms-$\eta$. Spin would always be oriented along the $z$ axis by the time it reaches the end location. Therefore, advancing on arms-$\delta$, spin would be precessing in a circular fashion with a net zero expectation energy, and is thus precluded from generating the dynamic phase. But at the values of $\eta=n\pi$, the total phase alternates between $0$ and $\pi$ on arms-$\delta$. For $\eta=2n\pi$, the denominator is always positive, and the device generates a total phase of $0$. For $\eta=(2n+1)\pi$, the denominator is always negative, and the total phase is $\pi$. Since the dynamic phase is always $0$, the total phase at $\eta=n\pi$ is also the geometric phase. For other values of $\eta$, the total phase takes on continuous values as a function of $\eta$ and $\delta$. Analysis above is focused only on the phases of arms-$\delta$. Phases on arms-$\eta$ for different $\eta$ values could, on the other hand, be found by prescribing $\delta=0$, details of which would be discussed later. For illustration, we refer to FIG.\ref{fig4} for $\eta=\pi,2\pi$ and observe the geometric phases on arms-$\delta$.\par
	\begin{figure}
		\centering
		\includegraphics[scale=0.5]{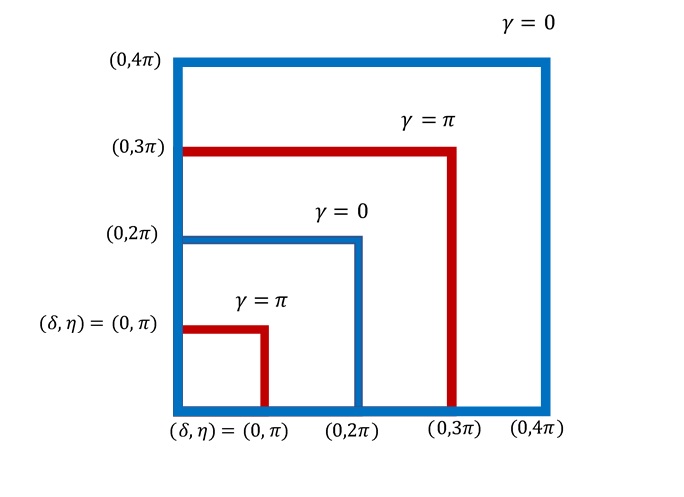}
		\caption{Schematic illustration of the effect of $\eta$ on the geometric phase $\gamma$ on arms $\delta$. The red and blue lines represent the paths with $\pi$ and $0$ geometric phases, respectively.}
		\label{fig4}
	\end{figure}
	In the case of partial entanglement, i.e. $p_{0}\neq p_{1}$
	\begin{align}
		\begin{split}
			\gamma=\tan&^{-1}{\frac{(p_{0}-p_{1})(\sin\delta\sin\eta)}{(\cos\delta+\cos\eta)\mp\sqrt{p_{0}p_{1}}\sin\eta\sin\delta}}\\
			&+2(\sin\eta)(p_{0}-p_{1})\delta\label{eq21}
		\end{split}
	\end{align}
	Like in the above, the dynamic phase can be eliminated by $\eta=n\pi$. Once again at these values, the total phase is discrete and alternates between $0$(for $\eta=2n\pi)$ and $\pi$(for $\eta =(2n+1)\pi)$. As before, the total phase at $\eta=n\pi$ is also the geometric phase. For other values of $\eta$, the total phase takes on continuous values as a function of $\eta$ and $\delta$. Note again that analysis here is focused only on the phases of arms-$\delta$.\par
	We will now revert to the case of maximum entanglement again. It was known that at maximal entanglement, the dynamic phase vanishes and the geometric phase takes on discrete values of $0$ and $\pi$. We would now study the exact locations on the QSR where the geometric phase switches its value. As a matter of fact, the positions of switching from $0$ to $\pi$ happens on arms-$\delta$. The exact location can be pinpointed by checking that the denominator of the geometric phase factor satisfies
	\begin{align}
		2(\cos\delta+\cos\eta)\mp (\sin\delta\sin\eta)>0\label{eq22}
	\end{align}

\begin{figure}
		\centering
		\includegraphics[scale=0.5]{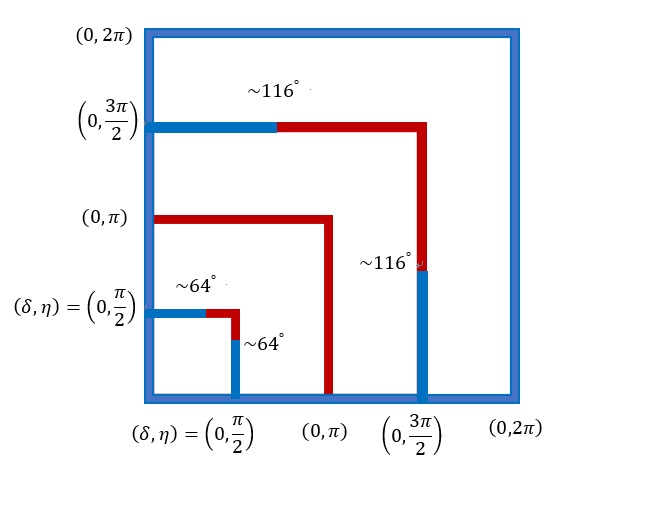}
		\caption{ Quantum square rings (QSR) of different sizes are superimposed for ease of inspection. The red and blue lines represent the paths with $\pi$ and $0$ geometric phases, respectively.}
		\label{fig5}
	\end{figure}

	The equation above shows that the answer would depend on the length of arms-$\eta$, i.e. the length of the arms before advancing into arms-$\delta$. For illustration, we chose arm lengths that correspond to $\eta=\frac{\pi}{2},\pi,\frac{3\pi}{2},2\pi$ as shown in FIG.\ref{fig5}. The device generates a $\gamma=0$ on arms $\eta$ at all times as indicated in blue. As the bipartite spin pair advances into arms-$\delta$, the geometric phase would switch to $\pi$ on locations as indicated by the red segments. At $\eta=2\pi$ though, no switching is possible and the geometric phase remains $0$ at all times. In the event of a zero denominator, the total phase $arg(\langle\psi(0)|\mathcal{U}|\psi(0)\rangle)=arg(a+ib)=\tan^{-1}{\frac{0}{0}}$ is undefined. The bipartite state at that juncture would have to either vanish or turn out orthogonal to the initial Bell states. 
	
\begin{table}\tiny
	\renewcommand{\tablename}{Table.}
	\centering
	
	\begin{tabular}{|c|c|c|c|c|}
		\hline
		\multicolumn{5}{|c|}{Non-Abelian: Non-adiabatic}\\
		\hline
		\multicolumn{5}{|l|}{$\ket{\phi(0)}=\sqrt{p_{0}}\ket{00}\pm\sqrt{p_{1}}\ket{11}$} \\ 
		\hline
		\multirow{2}{*}{~} & \multicolumn{2}{c|}{Dynamic phase} & \multicolumn{2}{c|}{Geometric phase}\\
		\cline{2-5}
		& $\delta=0$ & $\delta>0$ & $\delta=0$ & $\delta>0$\\
		\hline
		\makecell[c]{$(p_{0},p_{1}=(0,1))$ \\ $(p_{0},p_{1}=(1,0))$ \\ NO ENTG.} & 0 & 0 & 0 & 0\\
		\hline
		\makecell[c]{$p_{0}=p_{1}=\frac{1}{2}$\\ MAX. ENTG.} & 0 & 0 & 0 & 0\\
		\hline
		\makecell[c]{$p_{0}\neq p_{1}\neq 0$\\PARTIAL ENTG.} & 0 & 0 & 0 & 0\\
		\hline
		\multicolumn{5}{|c|}{}\\
		\hline
		\multicolumn{5}{|l|}{$\ket{\psi(0)}=\pm\sqrt{p_{0}}\ket{10}+\sqrt{p_{1}}\ket{01}$}\\
		\hline
		\multirow{2}{*}{~} & \multicolumn{2}{c|}{\makecell[c]{Dynamic phase\\$2(\sin\eta)(p_{0}-p_{1})\delta$}} & \multicolumn{2}{c|}{Geometric phase}\\
		\cline{2-5}
		& $\delta=0$ & $\delta>0$ & $\delta=0$ & $\delta>0$\\
		\hline
		\makecell[c]{$(p_{0},p_{1}=(0,1))$ \\ $(p_{0},p_{1}=(1,0))$ \\ NO ENTG.} & 0 & \makecell[l]{$0(\eta=2n\pi)$ \\ $0(\eta=(2n+1)\pi)$ \\ Continuous values as \\ $D=-(\mp2(\sin\eta)\delta)$} & 0 & \makecell[l]{Discrete $0(\eta=2n\pi)$ \\Discrete $\pi(\eta=(2n+1)\pi)$ \\ Continuous values as \\ $\gamma = \tan^{-1}{\frac{\mp(\sin\delta\sin\eta)}{\cos\delta+\cos\eta}}$\\ $\mp2(\sin\eta)\delta$}\\
		\hline
		\makecell[c]{$p_{0}=p_{1}=\frac{1}{2}$\\ MAX. ENTG.} & 0 & 0 & 0 & \makecell[l]{Discrete $0$ or $\pi$}\\
		\hline
		\makecell[c]{$p_{0}\neq p_{1}\neq 0$\\PARTIAL ENTG.} & 0 & \makecell[l]{$0(\eta=2n\pi)$ \\ $0(\eta=(2n+1)\pi)$ \\ Continuous values as \\ $D=-(2(\sin\eta)\delta)$} & 0 & \makecell[l]{Discrete $0(\eta=2n\pi)$ \\Discrete $\pi(\eta=(2n+1)\pi)$ \\ Continuous values \\ see Equation (21)}\\
		\hline
	\end{tabular}
\caption{Analysis of geometric phases for the SQR is tabulated according to the entanglement strength and the locations on arms-$\delta$.}\label{table1}
	
\end{table}
	
	Last is the particular situation of $\delta=0$ that corresponds to the point where spin-pair starts to take a right-angle bend into arms-$\delta$. As long as $\delta=0$, spin-pair is considered to reside in the $\eta$ regions of the arms only. And a quick inspection shows that $a(0,\eta)$ is positive throughout, which leads to the conclusion that the geometric phase on arms-$\eta$ is $0$ throughout, in spite of the entanglement strength. This is in fact indicated in FIG.\ref{fig4} and FIG.\ref{fig5} where arms-$\eta$ are painted blue to indicate a zero geometric phase throughout. This is indeed the case, barring the issues of singular points corresponding to $\cos\eta=-1$ which brings upon $\gamma=\tan^{-1}{\frac{0}{0}}$. At these points, the geometric phase is undefined. In terms of spin precession, the singular points correspond to spin making a rotation of $(2n+1)\pi$. The odd-pi quantum states of the spin-pair at this point would then be orthogonal to its initial Bell states. In terms of the dynamic phase, $\delta=0$ suppresses dynamic phases in spite of the entanglement strength. Table. I provides a summary of all the analysis that have been carried out for the geometric and dynamic phases corresponding to all the Bell states spin-pair traversing a non-Abelian QSR device.

	\section{Conclusion}
	We have explained in details how a non-Abelian system in the form of a QSR could be designed to generate and purify the Aharonov Casher phases into its geometric and dynamic components without elaborate experimental set ups. The device requires only an entangled-particle source to couple to a passive square ring. The Aharonov Casher phase is generated or annihilated as determined by the choice of the entanglement configuration. In the correct Bell states, the dynamic phase is eliminated outright at maximal entanglement. In the case of partial to no entanglement, dynamic phases are eliminated at $\eta=n\pi$. In all manners of elimination, the Aharonov Casher phase becomes discrete and fully geometric. This device could thus be useful for future experimental efforts to study the physics of discrete geometric phases. The continuous spectrum of the Aharonov Casher phase remains accessible though at partial to no entanglement, in which case, the continuous phases are non-geometric. At maximum entanglement, there is no possibility to access any continuous form of the geometric phase. In terms of discrete phases, the manner in which the phase switch from one discrete value to another varies according to the entanglement strength. At partial to no entanglement, switching occurs only at $(\delta,\eta)=(0,(2n+1)\pi)$. By contrast, at maximal entanglement, switching could take place anywhere on arms-$\delta$, i.e. any value of $(\delta,\eta)$. In summary, the device has been shown to generate continuous Aharonov Casher phases, annihilate dynamic phases, distill discrete geometric phases, and enable discrete phase switching at various locations, all within the simple construct of a square ring.
	\section{Acknowledgement}
	We would like to thank the Ministry of Science and Technology of Taiwan for supporting this work under Grant. No.: 110-2112-M-034-001-MY3.

	\bibliographystyle{apsrev4-2} 
	\bibliography{library} 
	
\end{document}